\documentclass[twocolumn,twoside,preprintnumbers,amsmath,amssymb,showkeys,%
showpacs]{revtex4}
\usepackage{graphicx}
\usepackage{dcolumn}
\usepackage{pslatex}

\graphicspath{{./Figures/}}

\newcommand{\etal}{\textit{\mbox{et al.\ }}}          
\newcommand{\ie}{\textit{\mbox{i.e.\ }}}              
\newcommand{\vs}{\textit{\mbox{vs.\ }}}               

\renewcommand{\Re}{\mbox{Re}\,}                       
\newcommand{\Tr}{\mbox{Tr}}                           
\newcommand{\bc}{\texttt{bc}}                         
\newcommand{\fc}{\texttt{fc}}                         
\newcommand{\Fig}[1]{Fig.~\ref{#1}}
\newcommand{\Tab}[1]{Table~\ref{#1}}

\newcommand{\Eq}[1]{Eq.~(\ref{#1})}

\newcommand{\beq}{\begin{equation}}
\newcommand{\eeq}{\end{equation}}
\newcommand{\bea}{\begin{eqnarray}}
\newcommand{\eea}{\end{eqnarray}}
\newcommand{\beas}{\begin{eqnarray*}}
\newcommand{\eeas}{\end{eqnarray*}}

\begin{document}
\preprint{HU-EP-06/32~~~LU-ITP 2006/017}
\title{{\Large 
Landau Gauge Gluon and Ghost Propagators from Lattice QCD
\footnote{Talk presented 
at the Workshop IRQCD '06, Rio de Janeiro, May 2006
by M. M{\"u}ller-Preussker.}
}}

\author{E.-M. Ilgenfritz, M. M{\"u}ller-Preussker, A. Sternbeck\footnote{
Address since Sept.~1, 2006: CSSM, School of Chemistry \& Physics, 
University of Adelaide, SA 5005, Australia.}}
\affiliation{Humboldt-Universit{\"a}t zu Berlin, Institut f{\"u}r Physik,
Newtonstr.~15, 12489~Berlin, Germany}

\author{A. Schiller}
\affiliation{Universit{\"a}t Leipzig, Institut f{\"u}r Theoretische Physik, 
Vor dem Hospitaltore~1, 04103~Leipzig, Germany}

\author{I. L. Bogolubsky}
\affiliation{Joint Institute for Nuclear Research, Dubna 141980, Russia}

\received{on October 24, 2006}

\begin{abstract}
We report on recent numerical computations of the Landau gauge gluon and 
ghost propagators as well as of the ghost-gluon-vertex function in pure 
$SU(3)$ Yang-Mills theory and in full QCD on the lattice. Special emphasis 
is paid to the low momentum region. In particular, we present new data
for the gluon propagator at momenta below 300 MeV. We also discuss
different systematic effects as there are finite-size, lattice
discretization and Gribov copy but also unquenching effects. 
A MOM-scheme running coupling $\alpha_s(q^2)$ based on the ghost-gluon
vertex is calculated and found to decrease for momenta below 550 MeV,
even though the renormalization constant of the vertex deviates only weakly
from being constant.


\end{abstract}

\maketitle

\section{INTRODUCTION}
\label{sec:intro}
With our contribution to this workshop we give an overview on 
recent lattice computations of the Landau gauge gluon and ghost 
propagators in quenched ($N_f=0$) and in full QCD ($N_f=2$). The full
analysis and additional results of related observables can be found
in the Ph.D.\ thesis of one of us \cite{Sternbeck:2006rd}.

For QCD being \textit{the} theory of the strong interaction a coherent
description of all hadronic features directly based on the dynamics of
confined quarks and gluons, 
given in terms of all propagators and vertex
functions of QCD, should be available. As reported by R.~Alkofer 
at this conference, they may serve as an input from first
principles for the Bethe-Salpeter and Faddeev equations and this opens a way
to a model independent phenomenology of nonperturbative phenomena.

Lattice computations of gauge-variant Green functions 
have attracted more and more interest in recent years, because the results 
can be directly confronted with studies of 
continuous Dyson-Schwinger equations (DSE). Both nonperturbative approaches 
have their own limitations. Whereas the lattice approach  
is affected by finite-size and discretization errors, the DSE approach
requires truncations of an infinite tower of equations and those truncations
are difficult to control. 
Therefore, a comparison of results is eventually able to provide more
confidence about the consistence of both approaches. 

Starting with the work by L.~von~Smekal \etal
\cite{vonSmekal:1997is,vonSmekal:1998}, DSE studies in recent years
\cite{Alkofer:2000wg,Bloch:2001wz,Fischer:2002eq,
  Fischer:2002hn,Fischer:2003zc,Fischer:2006vf,Pawlowski:2003hq} 
have shown evidence for an intertwined infrared power behavior of the 
gluon and ghost dressing functions
\bea \nonumber
Z(q^2) &\equiv& q^2 D(q^2) \propto (q^2)^{2\lambda}\,, \\
J(q^2) &\equiv& q^2 G(q^2) \propto (q^2)^{-\lambda}\,, 
\label{eq:infrared-behavior}
\eea
respectively, with the same value $\lambda\approx 0.59$
\cite{Lerche:2002ep,Zwanziger:2001kw}.
Thus, the gluon propagator $D(q^2)$ would be vanishing in the infrared 
in close connection with a diverging ghost propagator $G(q^2)$. This
infrared behavior is closely related to gluon and quark 
confinement in accordance with the Gribov-Zwanziger horizon condition
\cite{Zwanziger:1993dh,Gribov:1977wm} and the Kugo-Ojima criterion
\cite{Kugo:1979gm}. 

As a by-product, a nonperturbative determination of
the running coupling $\alpha_s(q^2)$ in a momentum subtraction (MOM)
scheme can be obtained. In fact, under the condition that the 
ghost-gluon-vertex renormalization function $Z_1(\mu^2)$ is finite 
and constant (see \cite{Taylor:1971ff,Marciano:1977su} and 
\cite{Cucchieri:2004sq} for a recent $SU(2)$ lattice study) the
corresponding running coupling is defined by
\beq
  \alpha_s(q^2) = \frac{g^2}{4\pi}~ Z(q^2)~ J^2(q^2)\,.
\label{eq:runcoupling}
\eeq
This together with relation (\ref{eq:infrared-behavior}) provides a
non-trivial fixed point of $\alpha_s$ in the infrared limit
\cite{Lerche:2002ep}, which was also proposed by D.~V.~Shirkov on the
basis of a perturbative analytic approach (see \cite{Shirkov:2002gw}
and references therein). 

For quenched $SU(2)$ extensive lattice investigations of the gluon and
ghost propagators in Landau gauge can be found in
\cite{Bloch:2003sk}. For $SU(3)$ lattice computations of the gluon
propagators were reported already in  
\cite{Leinweber:1998im,Leinweber:1998uu,Bonnet:2000kw,Bonnet:2001uh}.
Later, other groups focused on the ghost propagator 
\cite{Furui:2003jr,Furui:2004cx,Boucaud:2005ce}, too. In our study we
first paid special attention to the Gribov copy problem
\cite{Sternbeck:2005tk}. In the latter context we have also investigated 
spectral properties of the Faddeev-Popov operator \cite{Sternbeck:2005vs}. 
With increasing physical volume we have found the low-lying eigenmode spectrum 
becoming steeper at the lowest non-zero eigenvalues. This hopefully closes 
the gap from the trivial zero eigenvalues and leads to a non-vanishing 
spectral density at zero in the thermodynamic limit which is 
required for an infrared diverging ghost propagator. Moreover, in
\cite{Sternbeck:2005qj} we have reported on a first $SU(3)$ lattice computation 
of the ghost-gluon vertex at zero gluon momentum that lends confirmation (see 
below) for an almost constant ghost-gluon-vertex renormalization constant. 
This was shown earlier in \cite{Cucchieri:2004sq} for the case of $SU(2)$.

Also investigations for the gluon and ghost propagator in full QCD
have been reported \cite{Bowman:2004jm,Furui:2005bu,Furui:2005he}. 
We have extended our investigations to this case, too, 
using configurations generated with $N_f=2$ dynamical clover-improved Wilson 
fermions \cite{Ilgenfritz:2006gp}. 
The lattice field configurations have been provided to us by the QCDSF
collaboration \cite{Gockeler:2004rp,Gockeler:2005rv}. 
 
Finally, we only mention the analysis of some interrelated confinement
criteria: a check of reflection positivity violation by the gluon 
propagator \cite{Cucchieri:2006xi} and the computation of the Kugo-Ojima 
confinement parameter \cite{Furui:2006nm}. Our own checks of these
issues (see \cite{Sternbeck:2006rd}) have led to similar observations
and will be published elsewhere \cite{Sternbeck:2006cg}.

\section{LANDAU GAUGE, LATTICE GLUON AND GHOST PROPAGATORS} 
\label{sec:definitions}

The $SU(3)$ lattice gauge field configurations 
\mbox{$U=\{U_{x,\mu}\}$} 
have been generated by a standard (Hybrid) Monte Carlo algorithm
and then put into the Landau gauge 
by iteratively maximizing the gauge functional 
\beq
  F_{U}[g] = \frac{1}{4V}\sum_{x}\sum_{\mu=1}^{4}\Re\Tr \;{}^{g}U_{x,\mu}\,,
  \qquad 
  {}^{g}U_{x,\mu}=g_x\, U_{x,\mu}\,g^{\dagger}_{x+\hat{\mu}}
  \label{eq:functional}
\eeq
with $g_x \in SU(3)$. 
In general there are numerous local maxima (Gribov copies), 
each satisfying the lattice Landau gauge condition 
\beq
  (\partial_{\mu}{}^{g}\!\!A_{\mu})(x) \equiv \sum_{\mu} 
  \left( {}^{g}\!\!A_{\mu}( x+ \hat\mu/2)-{}^{g}\!\!A_{\mu}( x- \hat\mu/2) 
  \right) = 0
  \label{eq:transcondition}
\eeq 
for the gauge transformed lattice potential 
\beq 
  {}^{g}\!\!A_\mu(x+\hat{\mu}/2) = \frac{1}{2i}\left(^{g} U_{x,\mu} -
                    \ ^{g} U^{\dagger}_{x,\mu}\right)\Big|_{\rm
                    traceless}.
\label{eq:A-definition}
\eeq
To explore to what extent this ambiguity has a significant influence on gauge
dependent observables, for lattice sizes up to $24^4$ we have gauge fixed each 
thermalized configuration a certain number of times (several tens depending 
on the lattice size and coupling constant) employing the \emph{over-relaxation} 
algorithm and starting always from purely random gauge copies. 
Then for each configuration $U$, we have selected the
first (\fc{}) and the best (\bc{}) gauge copy 
(best with respect to the functional value) for subsequent measurements. 
For details we refer 
to \cite{Sternbeck:2005tk}. On lattices sizes larger than $24^4$ we have 
restricted ourselves only to \emph{one} copy per thermalized configuration 
(only \fc{} copies). On those lattices we have
applied also the \emph{Fourier accelerated} method for fixing $U$ to the
Landau gauge.

The momentum space gluon propagator $D^{ab}_{\mu\nu}(q^2)$ is 
the correlator of two Fourier transforms $\widetilde{A}^a_{\mu}(k)$ 
of the lattice potential ${}^{g}\!\!A_\mu(x+\hat{\mu}/2)$ (the symbol 
${}^{g}$ is dropped from now on) 
\bea \nonumber
D^{ab}_{\mu\nu}(q^2) &=& 
\left\langle \widetilde{A}^a_{\mu}(k)\widetilde{A}^b_{\nu}(-k) \right\rangle \\ 
&=& \delta^{ab} \left( \delta_{\mu\nu} - \frac{q_{\mu}~q_{\nu}}{q^2} \right) 
    D(q^2)\;,
\label{eq:D-definition}
\eea
where $q$ denotes the ``physical'' momentum
\beq
  q_{\mu}(k_{\mu}) = \frac{2}{a} \sin\left(\frac{\pi
      k_{\mu}}{L_{\mu}}\right)
\label{eq:p-definition}
\eeq
related to the integer valued lattice momentum 
$k_{\mu}\in \left(-L_{\mu}/2, L_{\mu}/2\,\right]$ for the linear 
lattice extensions $L_{\mu}, \mu=1,\ldots, 4$. According to 
Ref.~\cite{Leinweber:1998uu}, a subset of 
admissible lattice momenta 
$k$ has been chosen for the final analysis of the gluon (and ghost) propagator, 
although the Fast Fourier Transform algorithm easily provides us with all
lattice momenta. 

The ghost propagator is derived from the Faddeev-Popov (F-P) operator,
the Hessian with respect to $g_x$ of the gauge functional given 
in \Eq{eq:functional}.
It can be written in terms of the (gauge fixed) link variables $U_{x,\mu}$ 
as
\bea
  M^{ab}_{xy} & = & \sum_{\mu} A^{ab}_{x,\mu}\,\delta_{x,y}
  - B^{ab}_{x,\mu}\,\delta_{x+\hat{\mu},y}
  - C^{ab}_{x,\mu}\,\delta_{x-\hat{\mu},y}\quad
  \label{eq:FPoperator}
\eea
with\vspace{-0.78cm}
\beas
  A^{ab}_{x,\mu} &=& \phantom{2\cdot\ } \Re\Tr\left[
    \{T^a,T^b\}(U_{x,\mu}+U_{x-\hat{\mu},\mu}) \right],\\
  B^{ab}_{x,\mu} &=& 2\cdot\Re\Tr\left[ T^bT^a\, U_{x,\mu}\right],\\
  C^{ab}_{x,\mu} &=& 2\cdot\Re\Tr\left[ T^aT^b\, U_{x-\hat{\mu},\mu}\right]
\eeas
and $~T^a,~a=1,\ldots,8~$ being the (hermitian) generators of the
$~\mathfrak{su}(3)~$ Lie algebra satisfying $~\Tr~[T^aT^b]~=~\delta^{ab}/2$.
The ghost propagator is then determined by inverting the F-P operator $~M$ 
\bea
  G^{ab}(q) &=& \frac{1}{V} \sum_{x,y} \left\langle {\rm e}^{-2\pi i\,k
      \cdot (x-y)} [M^{-1}]^{ab}_{xy} \right\rangle_U= \delta^{ab} G(q^2)\,.
      \qquad
\label{eq:Def-ghost}
\eea
Following Refs.~\cite{Suman:1995zg,Cucchieri:1997dx} we have used the 
conjugate gradient (CG) algorithm to invert $M$ on a plane wave $\vec{
\psi}_c$ with color and position components \mbox{$\psi^a_c(x) = \delta^{ac}
\exp (2\pi i\,k\!\cdot\! x)$}. In fact, we applied a pre-conditioned
CG algorithm (PCG) to solve $M^{ab}_{xy}\phi^{b}(y)=\psi^a_c(x)$. As 
the pre-conditioning matrix we used the inverse Laplacian operator 
$\Delta^{-1}$ with a diagonal color substructure. This has significantly 
reduced the required amount of computing time (for details see 
\cite{Sternbeck:2005tk}). 

\section{PROPAGATORS: QUENCHED AND FULL QCD RESULTS}

\label{sec:propagators}

\begin{figure*}[htbp]
\begin{center}
\includegraphics[width=7.6cm
]{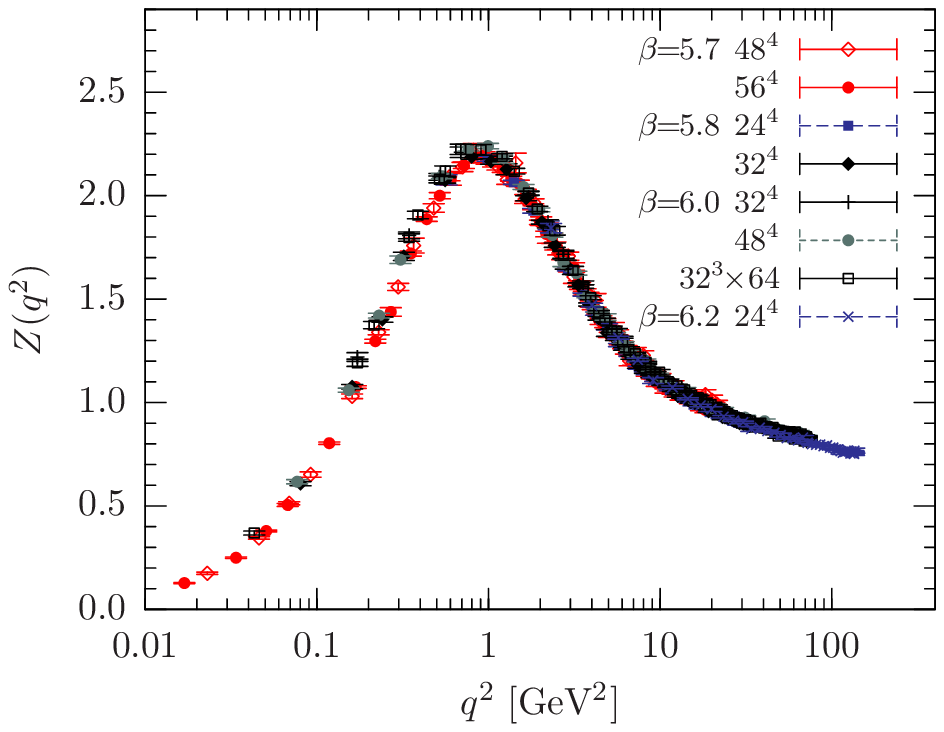} 
\hspace*{2.0cm}
\includegraphics[width=7.6cm
]{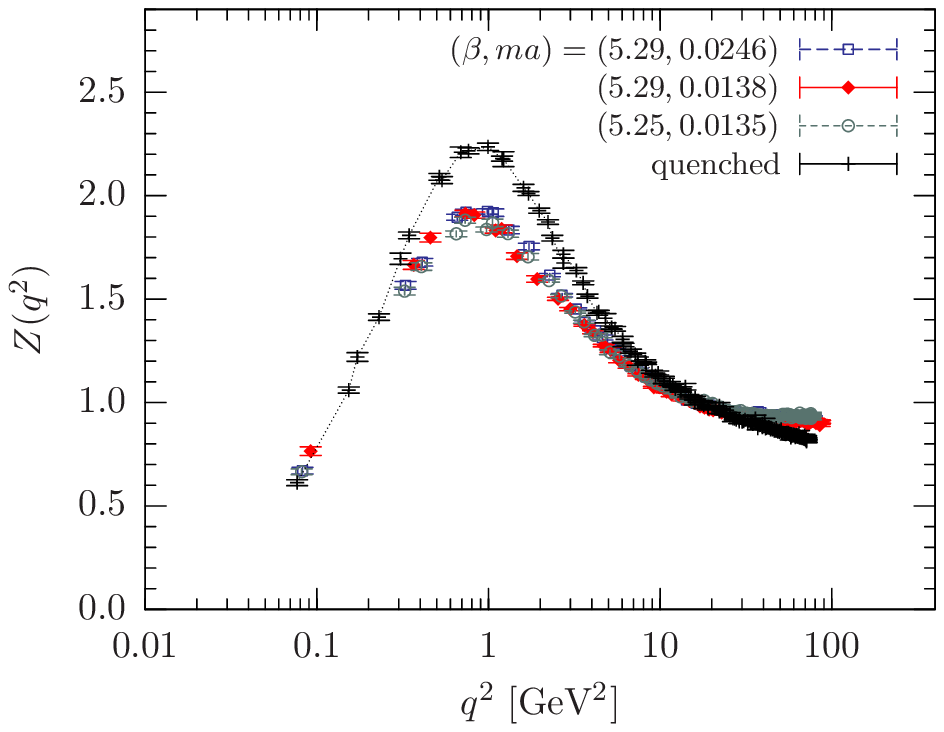} 
\caption{
The dressing functions for the gluon propagator 
$Z(q^2) \equiv q^2 D(q^2)$ \vs $q^2$
for quenched QCD (l.h.s.) and full QCD (r.h.s.), 
both measured on \fc{} gauge copies. To illustrate the unquenching
effect some quenched QCD data ($\beta=6.0$, $32^4$ and $48^4$) are shown  
in the right figure, too.
}
\label{fig:gluon_dress}
\end{center}
\end{figure*}
\begin{figure*}[htbp]
\begin{center}
\includegraphics[width=7.5cm
]{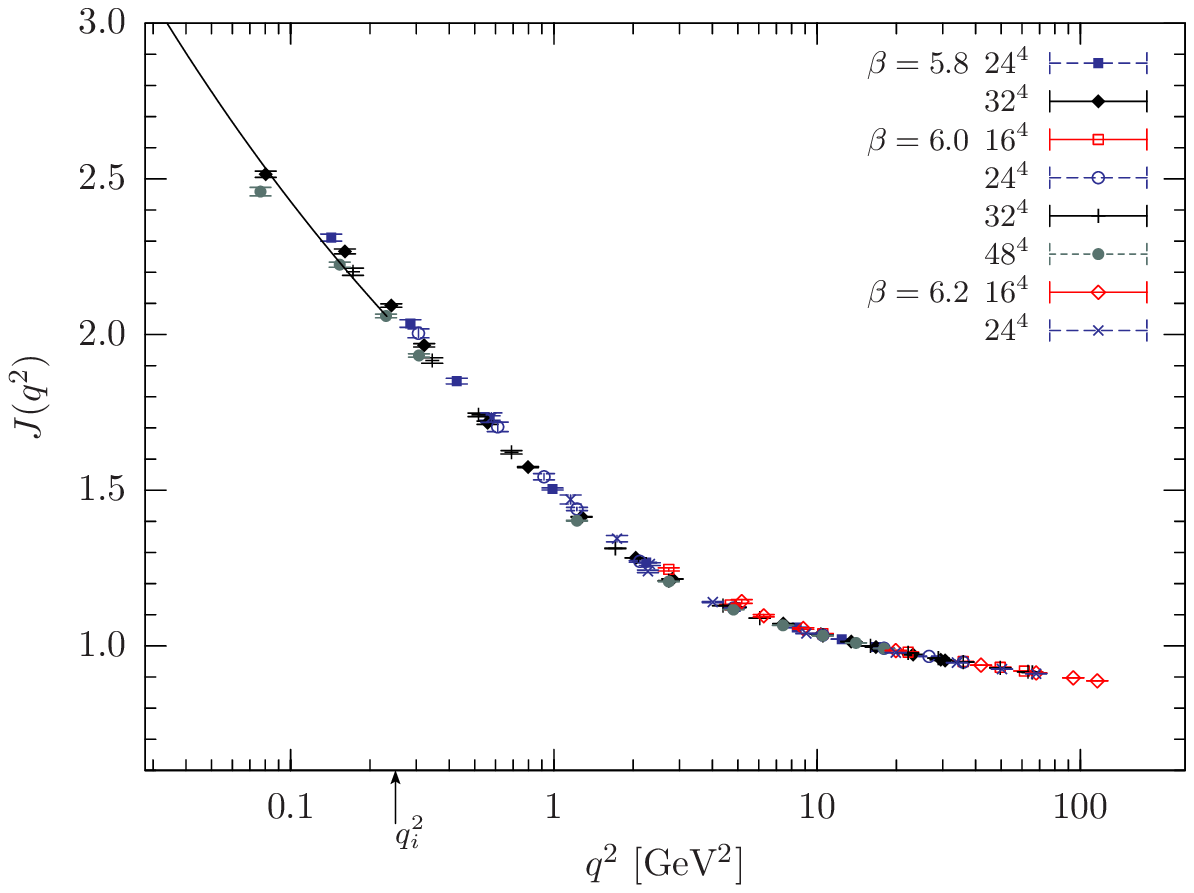} 
\hspace*{2.0cm}
\includegraphics[width=7.5cm
]{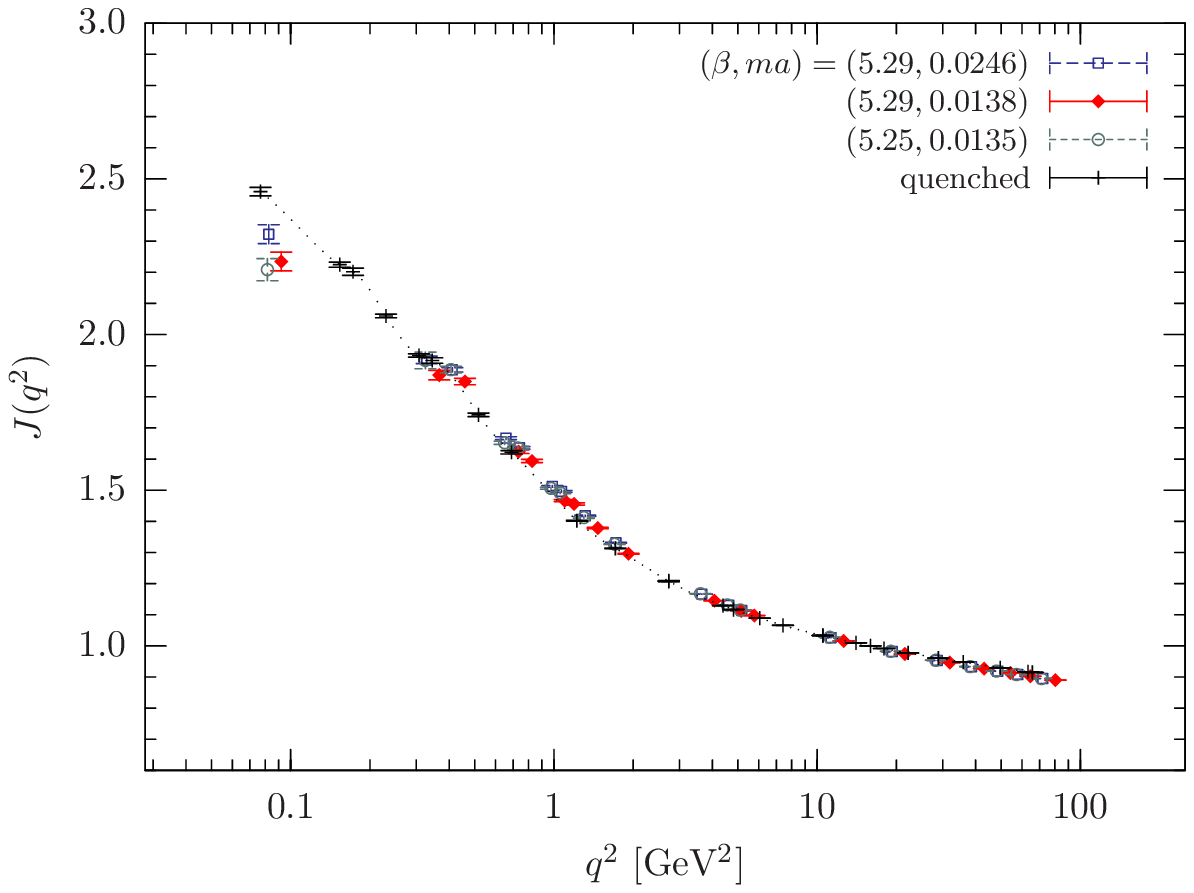} 
\caption{
As \Fig{fig:gluon_dress}, here for the dressing function
of the ghost propagator $J(q^2) \equiv q^2 G(q^2)$. The line (l.h.s.)
represents a fit to the data for momenta lower than $q_i^2$ using the
power law given in relation (\ref{eq:infrared-behavior}).} 
\label{fig:ghost_dress}
\end{center}
\end{figure*}
\begin{figure}[htbp]
\begin{center}
\includegraphics[width=7.5cm
]{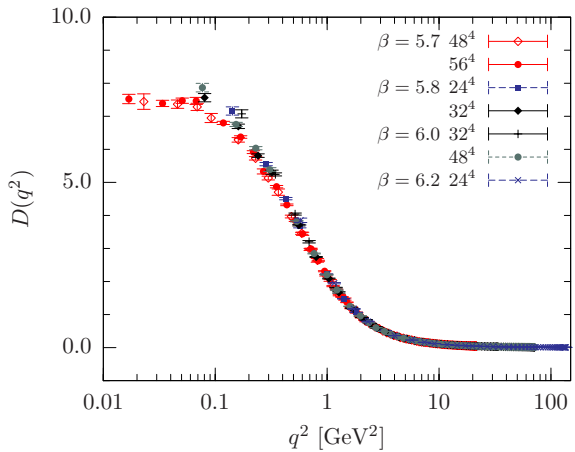} 
\caption{
The gluon propagator $~ D(q^2)~$ \vs $q^2$
for quenched QCD, measured on \fc{} gauge copies. 
}
\vspace*{-0.5cm}
\label{fig:gluon_prop}
\end{center}
\end{figure}

For the pure $SU(3)$ Yang-Mills case we have generated the gauge fields
using the standard Wilson plaquette action for bare coupling values 
$\beta=6/g_0^2 =\{5.7, 5.8, 6.0, 6.2\}$. Studying very large lattices (up to
lattice size $56^4$) at the smallest possible $\beta$ value we tried to  
probe the infrared limit, but with the reservation that the results can
be affected by lattice discretization artifacts. To match our
lattice results to physical units we used the parametrization of the
lattice spacing $a$ depending on $\beta$ as determined in
\cite{Necco:2001xg}. For the full QCD case we have used 
Hybrid Monte Carlo generated QCDSF
gauge field configurations produced with
$N_f=2$ dynamical flavors of clover-improved
Wilson fermions. The relevant parameter sets 
\cite{Gockeler:2004rp,Gockeler:2005rv} are listed in 
\Tab{tab:stat}. As lattice size we used always $24^3\times 48$. Note that 
the asymmetry demands some caution in the very infrared.

The gluon dressing function $Z(q^2)$ and the ghost one $J(q^2)$ 
are shown versus $q^2$ in physical units in \Fig{fig:gluon_dress} and
\Fig{fig:ghost_dress}, respectively, always for the \fc{} copies only. 
The data represent the current status of our quenched and full QCD 
calculations. In particular, for the quenched case we have included
very new results for the gluon propagator obtained at $\beta=5.7$ using
the lattice sizes $48^4$ and $56^4$. All dressing functions have been
renormalized separately for each $\beta$ such that they equal 
unity at $q = 4~\text{GeV}$.

At a first glance our data for the gluon dressing function seem to be 
qualitatively in agreement with an infrared suppression, whereas the ghost
dressing function looks compatible with an infrared singularity. 
The unquenching effect is clearly visible for the gluon propagator,
whereas the ghost propagator is almost unaffected by including the fermionic 
feed-back in the functional measure of the gluons. This does not come 
unexpected since the ghost fields do not directly couple to the fermion 
fields. The non-perturbative peak of the gluon dressing function 
at $q \simeq 1~\mbox{GeV}$ becomes softer as the quark mass is decreasing.
It should be recalled that the gluon peak is wholly removed when
center vortices are removed from the gluon field~\cite{Langfeld:2001cz}.
The corresponding effect of dynamical quarks has been observed also 
in other lattice computations of the gluon propagator using dynamical 
AsqTad improved staggered quarks \cite{Bowman:2004jm} 
and is expected from studies of the unquenched ghost 
and gluon propagators within the DSE approach 
\cite{Fischer:2005wx,Fischer:2005en,Fischer:2003rp}.
We refer also to lattice studies with dynamical Kogut-Susskind
and Wilson fermions reported in \cite{Furui:2005bu,Furui:2005he} and
to the contribution given by S.~Furui during this meeting 
\cite{Furui:2006nm}.

For the gluon as well as for the ghost dressing functions in the quenched 
case, with somewhat improved data in comparison to \cite{Sternbeck:2005tk},
we have tried fits of a power-like infrared behavior 
(see \cite{Sternbeck:2006rd}). 
For the ghost dressing function the result is indicated in the l.h.s.\ of 
\Fig{fig:ghost_dress}. The resulting exponent turned out to be quite 
stable against variations of the upper end of the fit interval $q_i^2$. 
We have found $\lambda=0.20(1)$, \ie much smaller than expected from the 
DSE approach. A similar observation was made for the gluon exponent.    
However, in order to check whether the gluon propagator really vanishes 
in the infrared --- in accordance with the Gribov-Zwanziger horizon
condition --- we show the gluon propagator itself in \Fig{fig:gluon_prop}. 
With our new quenched $SU(3)$ data obtained at $\beta=5.7$ on a $56^4$ lattice 
we now find a first indication for having reached a maximum, opening the 
opportunity for a decreasing gluon propagator for even lower momenta. 
Our results do not seem to expose dramatic finite-size effects,  
but obviously one needs much larger lattices in order to reach momenta 
well below $100~\text{MeV}$. Therefore, it is premature to attempt fits 
of the infrared exponents with the hope to reproduce 
the DSE results (see \Eq{eq:infrared-behavior}). This
certainly also holds for the ghost propagator, \ie the $\lambda$ value
quoted above cannot be taken seriously.

\begin{table}[t]
\centering
\caption{
The parameter sets $\beta$, $\kappa$ etc.\ for the configurations used 
in our full QCD investigation (courtesy of the QCDSF collaboration).
}
\begin{tabular}%
{@{\quad}c@{\quad}c@{\quad}c@{\quad}c@{\quad}c@{\quad}c@{\quad}}
\hline\hline\rule{0pt}{2.5ex}
$\beta$ & $\kappa$ & $\kappa_c$ & $ma$ & $a [\mbox{GeV}^{-1}]$ &\#~\mbox{conf} 
\\
\hline\rule{0pt}{2.5ex}
5.29 & 0.13550 & 0.136410\,(09) & 0.0246 & 2.197 & 60 \\
5.29 & 0.13590 & 0.136410\,(09) & 0.0138 & 2.324 & 55 \\
5.25 & 0.13575 & 0.136250\,(07) & 0.0135 & 2.183 & 60 \\
\hline\hline
\end{tabular}
\label{tab:stat}
\end{table}

\section{GHOST-GLUON-VERTEX FUNCTION AND THE RUNNING COUPLING}
\label{sec:coupling}

\begin{figure*}[htbp]
\begin{center} 
\includegraphics[width=7.0cm
]{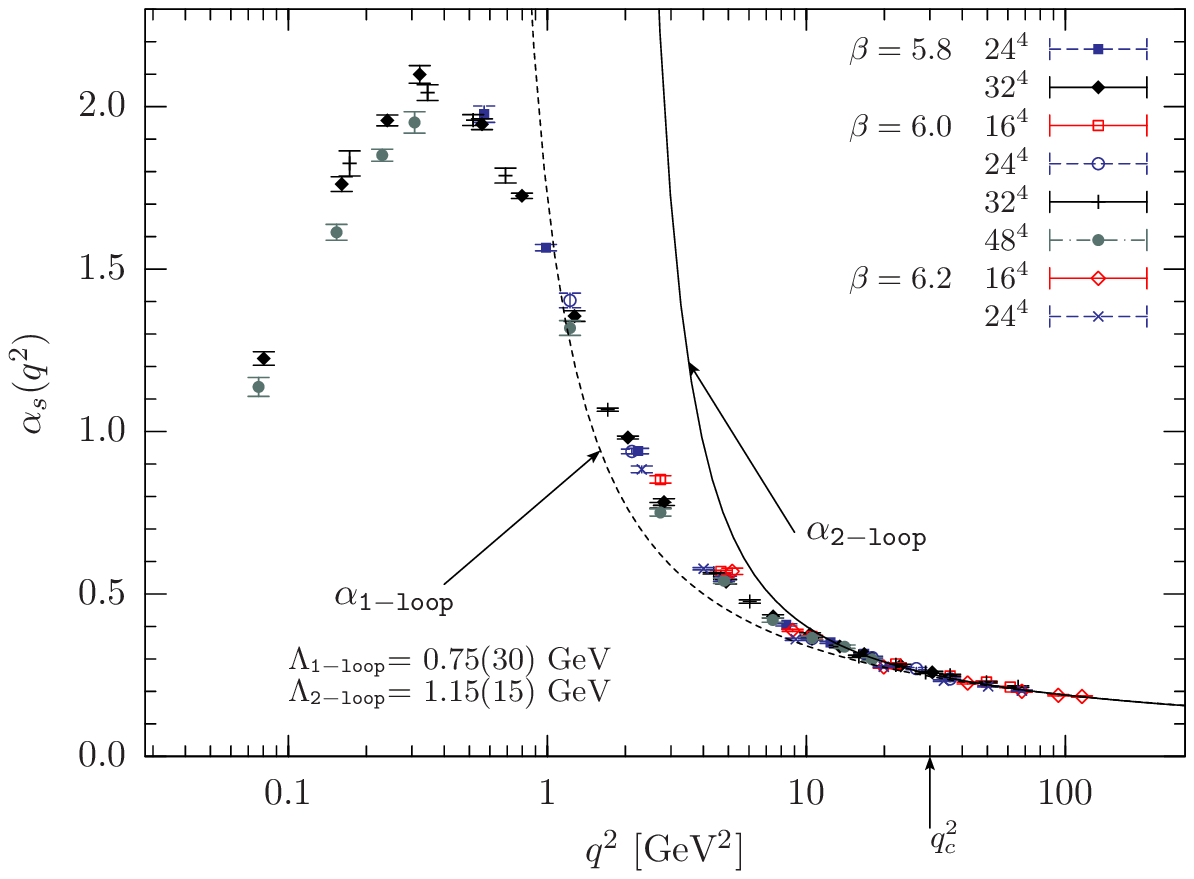} 
\hspace*{2.4cm}
\includegraphics[width=7.0cm
]{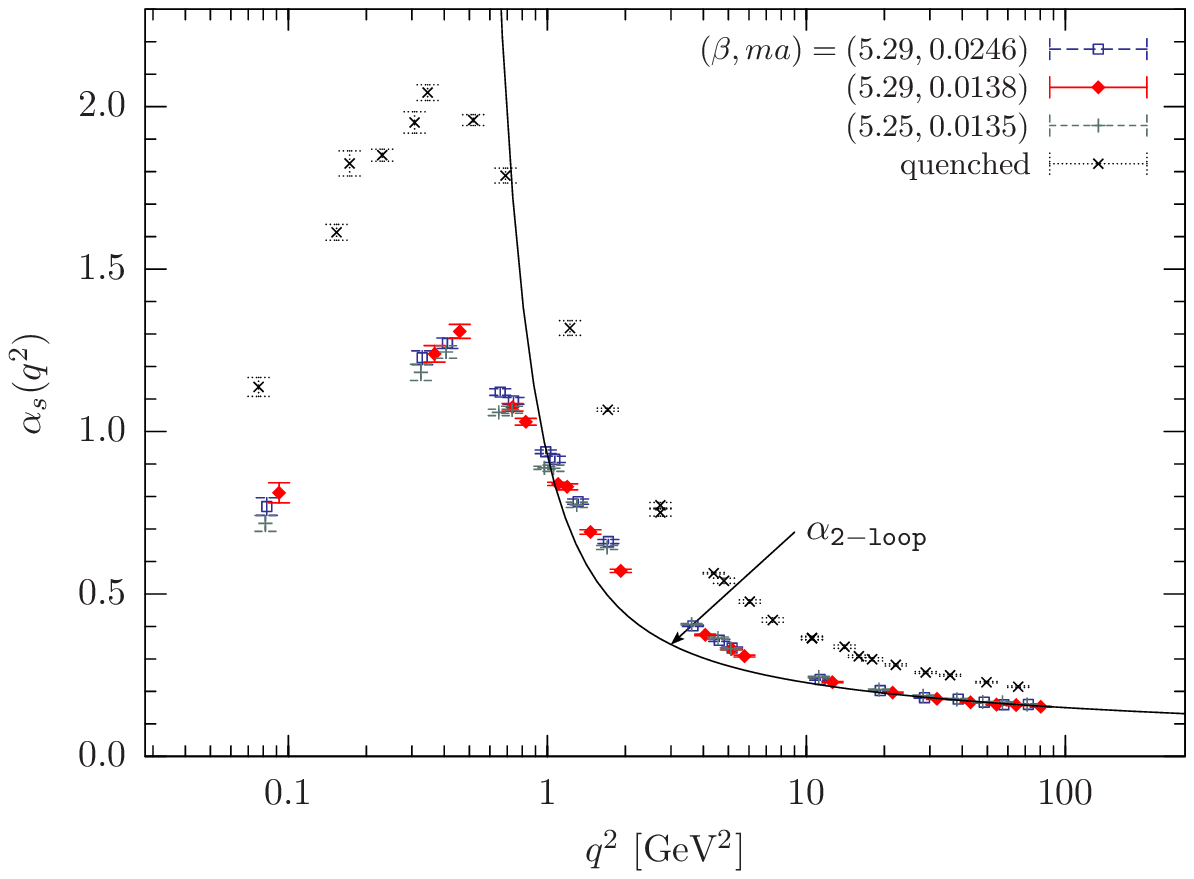}
\caption{
The momentum dependence of the running coupling $\alpha_{s}(q^2)$ 
for quenched QCD (l.h.s.) and full QCD (r.h.s.)i, measured on \fc{} 
gauge copies. 
For comparison, on the right hand side selected quenched QCD data 
for $\beta=6.0, ~32^4$ are shown. 
1- and 2-loop fits to $\alpha_s(q^2)$ are drawn with dashed and
solid lines, respectively. 
} 
\label{fig:alpha}
\end{center}
\end{figure*}
 
In \Fig{fig:alpha} we show the running coupling according to 
\Eq{eq:runcoupling}. Surprisingly, the coupling is seen to
decrease below $q^2 \simeq 0.3~\mathrm{GeV}^2$ rather than to approach 
the predicted non-trivial infrared fix-point monotonously from below.
The same happens as well for the quenched as for the unquenched 
case, irrespective of the clearly visible unquenching effects.
Whether strong finite-size effects in the ghost propagator may 
change the present tendency remains to be seen in future. From 
our checks of finite-size effects (see below) we cannot 
derive arguments that this will be the case. 
It is worth mentioning here that lattice 
computations of the running coupling from other vertex functions have 
provided quite similar results (see \cite{Boucaud:1998bq,Boucaud:2002fx} for 
the three-gluon vertex and \cite{Skullerud:2002ge} for the quark-gluon 
vertex). 

\begin{figure*}[htbp]
\begin{center}
\includegraphics[width=6.2cm
]{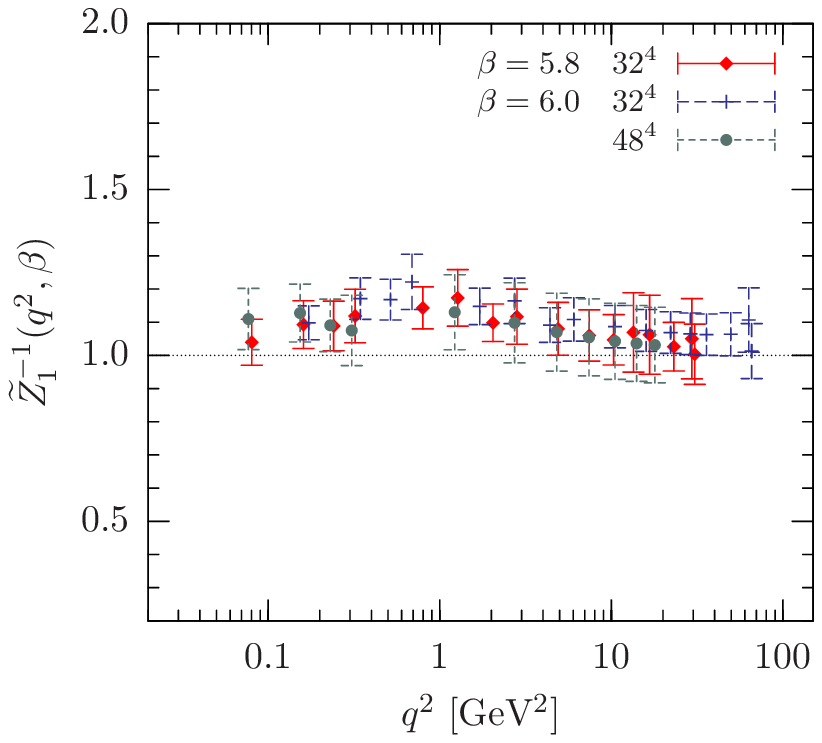}  
\hspace*{2.4cm}
\includegraphics[width=6.2cm
]{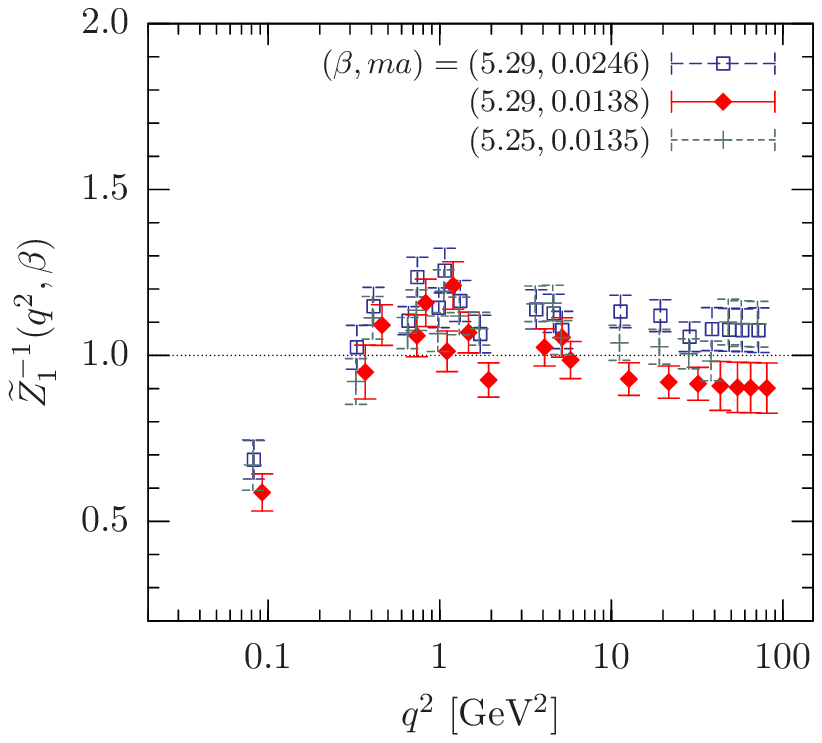}
\caption{
The inverse ghost-gluon-vertex renormalization function $Z_1^{-1}(q^2)$, 
measured on \fc{} gauge copies 
for quenched QCD (l.h.s.) and full QCD (r.h.s.). 
}
\label{fig:z1}
\end{center}
\end{figure*}

The definition of the running coupling relies on the assumption of a constant 
ghost-gluon-vertex renormalization function $Z_1(q^2)$. 
A recent lattice investigation of this function defined at vanishing 
gluon momentum for the $SU(2)$ case 
\cite{Cucchieri:2004sq} supports that \mbox{$Z_1(q^2)\approx 1$} at least 
for momenta larger than 1~GeV. We have performed an analogous study for 
$Z_1(q^2)$ in the case of $SU(3)$ gluodynamics and for full QCD. 
Our results are presented in \Fig{fig:z1}. 
There is a slight variation visible in the interval 
$0.3~\mbox{GeV}^2 \le q^2 \le 5~\mbox{GeV}^2$. However,
this weak deviation from being constant would not have a dramatic 
influence on the running coupling. 

The left-most two data points falling below unity on the r.h.s.\ of 
\Fig{fig:z1} correspond to the lowest on-axis momentum on the (asymmetric) 
$24^3\times48$ lattice. We believe that this deviation is due to
the asymmetry of the lattice, because we have seen a 
similar effect (not shown) in our quenched data, too. Simulations on larger 
(symmetric) lattices will enable us to sharpen our conclusions within 
the near future.

\section{SYSTEMATIC LATTICE AND GRIBOV COPY EFFECTS}
\label{sec:syseffects}

\begin{figure*}[htbp]
\begin{center}
\includegraphics[width=15cm,height=8.5cm]{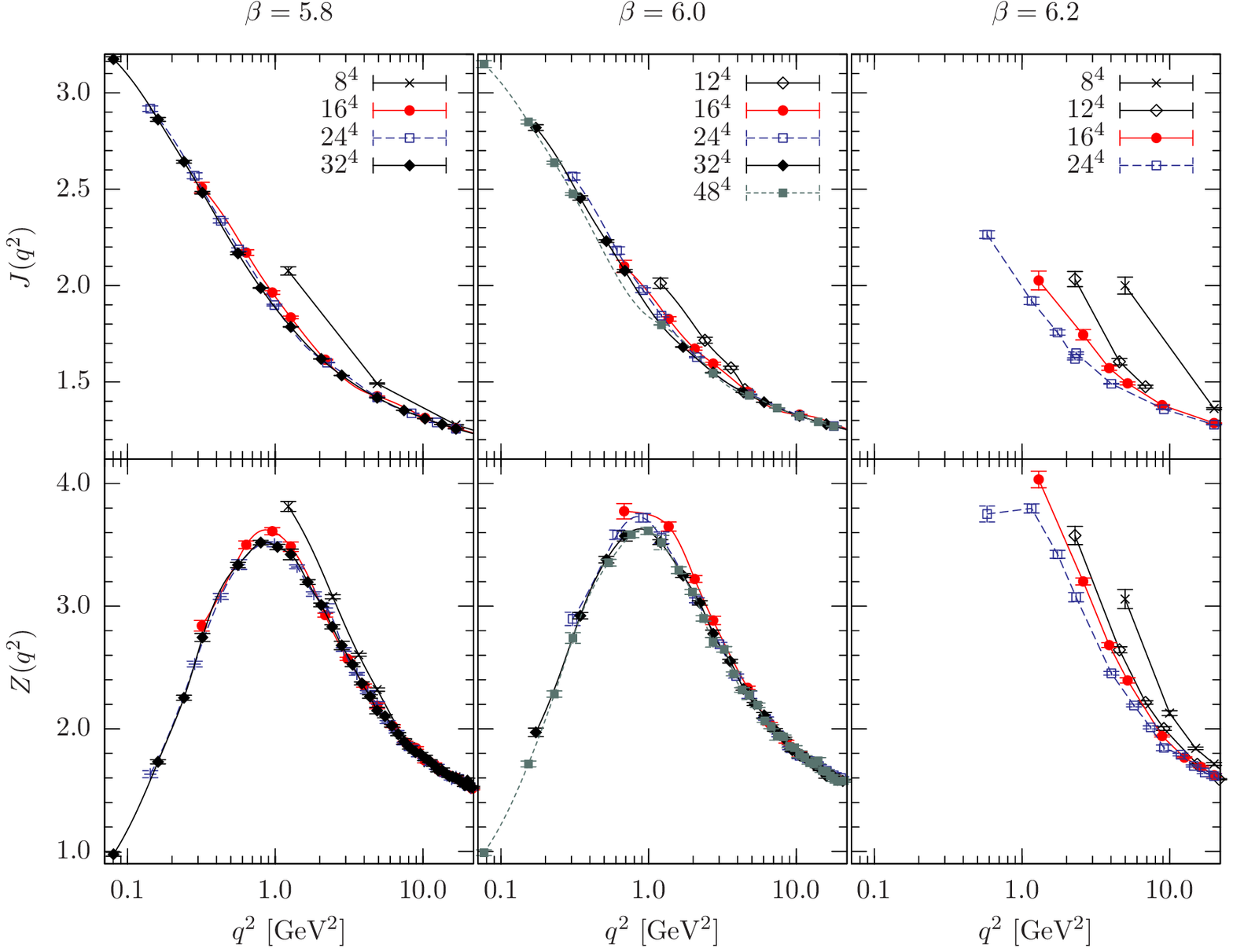} 
\caption{
Demonstration of finite-size effects of the dressing functions
for the ghost (upper row) and for the gluon (lower row) in quenched QCD. 
}
\label{fig:fs_effects}
\end{center}
\end{figure*}
It is a common problem of all lattice computations that one has to check
the sensitivity of the results with respect to finite-size effects, 
to the boundary conditions adopted, to an eventual lattice asymmetry, 
and to the given lattice discretization.
We discuss three of these checks in the following. 
Note that we have employed some cuts on the list of momenta used 
in order to minimize effects of the lattice discretization 
from the beginning (see \cite{Leinweber:1998uu},\cite{Sternbeck:2006rd}).

In \Fig{fig:fs_effects} we demonstrate how the finite lattice size can
influence the propagator results. For several fixed $\beta$ values, \ie
for fixed lattice spacings we check the dependence on the varying volume.
For smallest lattices (at smallest lattice spacings for $\beta=6.2$) we
find very strong finite-size effects in both cases as one has expected.
But our results show that at larger volumes (at lower $\beta$ values)
the lattice size dependence becomes less dramatic.

\begin{figure*}[htbp]
\begin{center}
\includegraphics[width=6.2cm
]{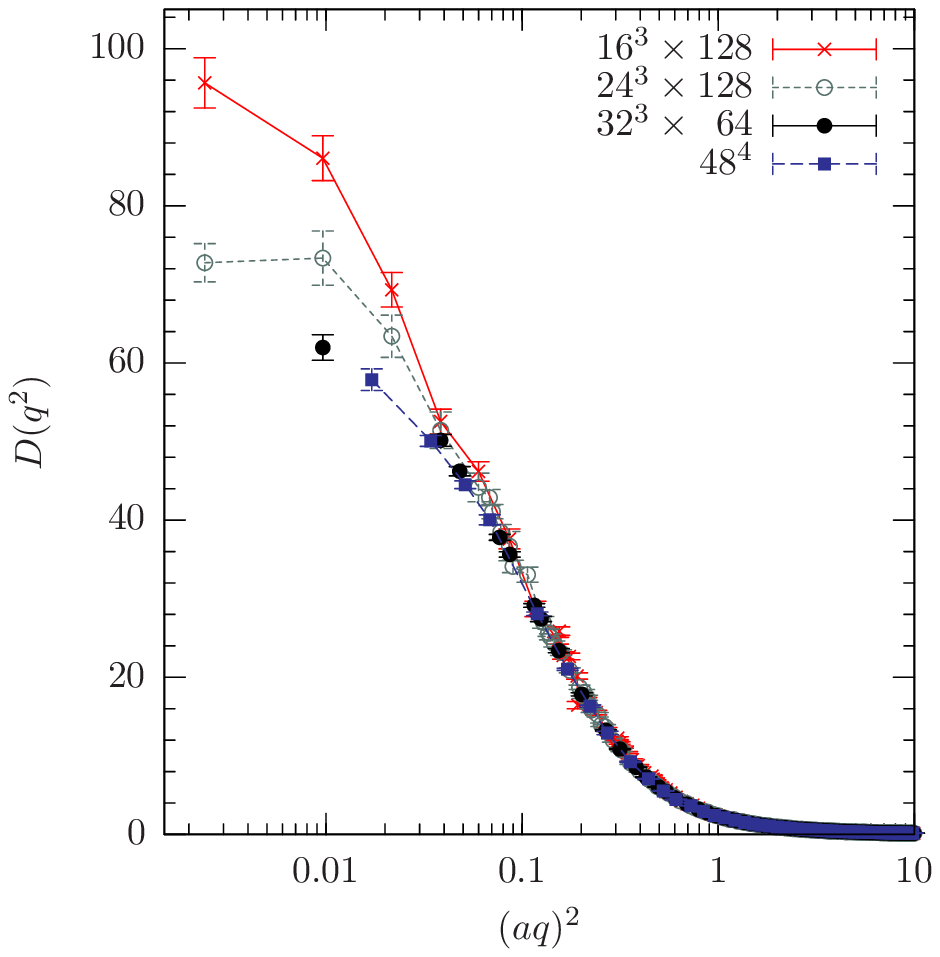} 
\hspace*{2.0cm}
\includegraphics[width=6.2cm
]{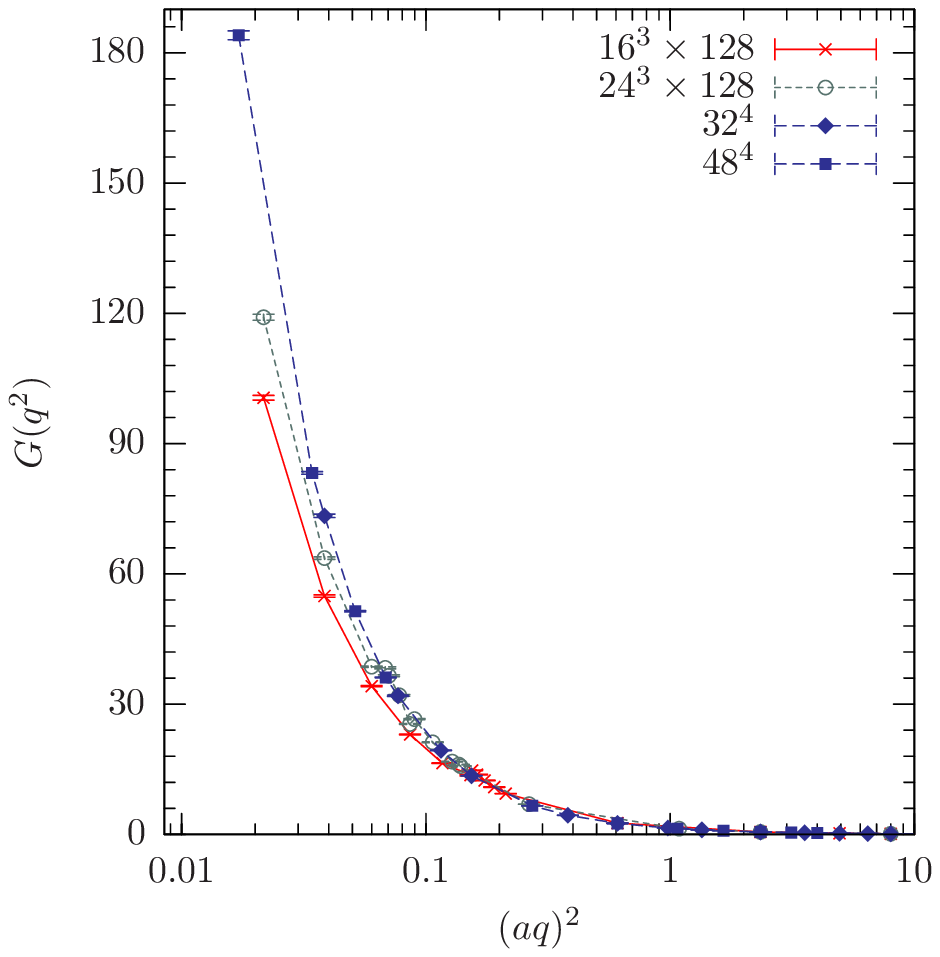}
\caption{
Lattice asymmetry effects for the gluon (l.h.s.) and ghost (r.h.s.)
propagators, respectively, in quenched QCD ($\beta=6.0$, only \fc{} 
gauge copies). Note that we have dropped data points at even lower 
momentum (on-axis in the elongated time direction). Our intention 
here is to concentrate on the momentum range where we can compare 
with data obtained on symmetric lattices.
}
\label{fig:asym_effects}
\end{center}
\end{figure*}

As a part of the project we also performed simulations at
$\beta=6.0$ using various asymmetric lattices. In \Fig{fig:asym_effects} 
we show gluon (l.h.s.) and ghost (r.h.s.) dressing function data 
obtained for lattice sizes $16^3\times128$ and $24^3\times128$ 
in comparison with the symmetric lattice $48^4$. 
We see large systematic effects at low momenta due to
the asymmetry, in particular, for the lowest on-axis momenta 
along the elongated time direction. For a more detailed study 
of this effect see \cite{Sternbeck:2006rd}. Similar observations have been
made from 3d $SU(2)$ investigations in \cite{Cucchieri:2006za}. 
At this workshop P. Silva has reported on the possibility to use 
the asymmetry effect for an extrapolation towards the large volume limit 
\cite{Silva:2005hb,Oliveira:2006yw}. It remains to be seen, whether such
an extrapolation really gives stable results. 

\begin{figure*}[htbp]
\begin{center}
\includegraphics[width=12.0cm,height=5.5cm]{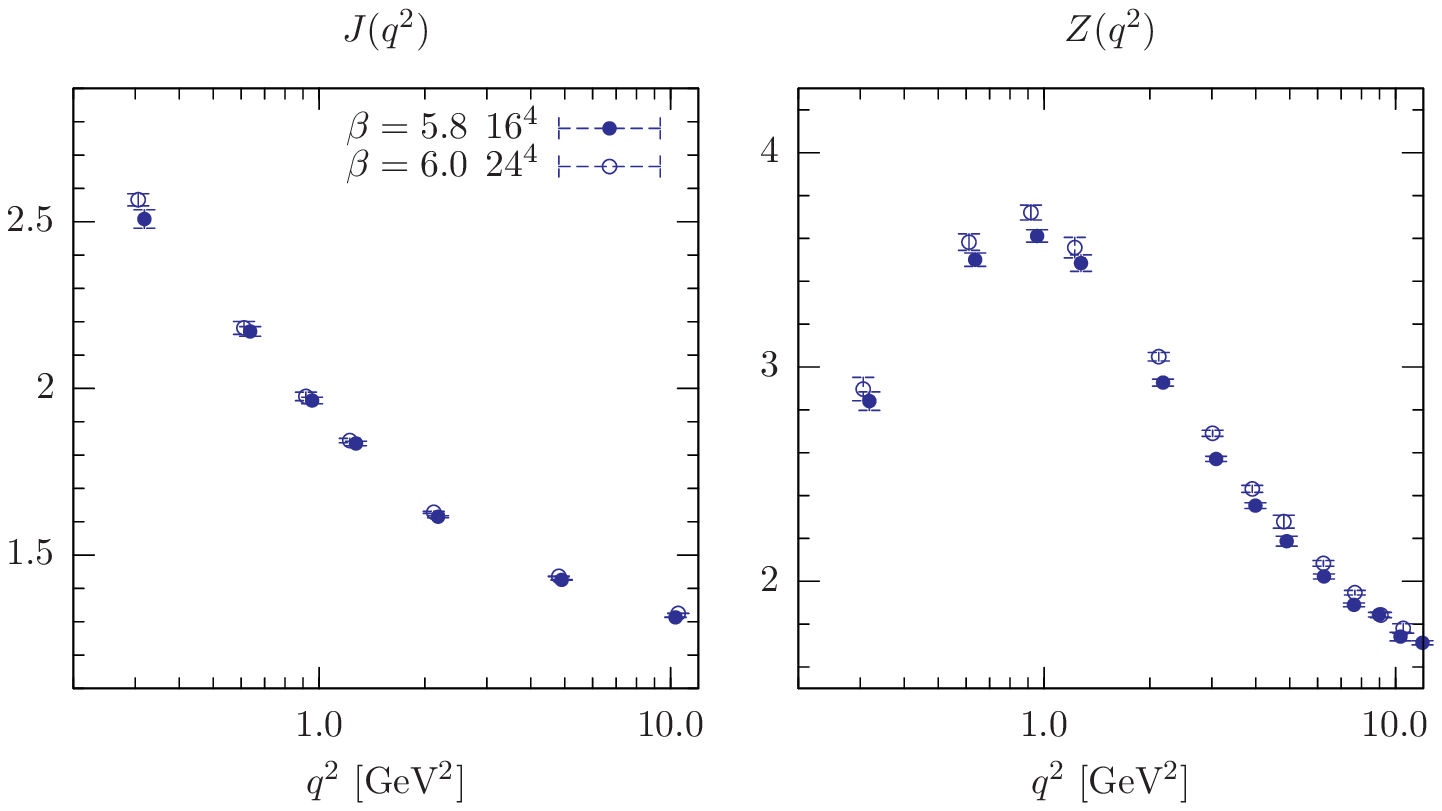} 
\caption{
Lattice discretization effects of the ghost (l.h.s.) and gluon
(r.h.s.) dressing functions, respectively, in the quenched case. 
Both dressing functions are shown for two cases ($\beta=5.8, 16^4$ 
and $\beta=6.0, 24^4$) corresponding to approximately the same lattice 
volume $\simeq (2.2~\mathrm{fm})^4$. }
\label{fig:latdis_effects}
\end{center}
\end{figure*}

Next, in \Fig{fig:latdis_effects} we present a check of the dependence 
on the lattice discretization for the ghost and the gluon dressing
functions in quenched QCD. Whereas the lattice volumes are approximately
the same, the lattice spacing changes from $~a \simeq 0.136~\mathrm{fm}~
(\beta=5.8)$ to $~a \simeq 0.093~\mathrm{fm}~(\beta=6.0)$. 
The comparison shows that
lattice discretization effects are almost negligible (for the ghost) 
or quite moderate (for the gluon) in the given momentum
interval for the set of (preselected) lattice momenta.  

\begin{figure*}[htbp]
\begin{center}
\includegraphics[width=7.0cm
]{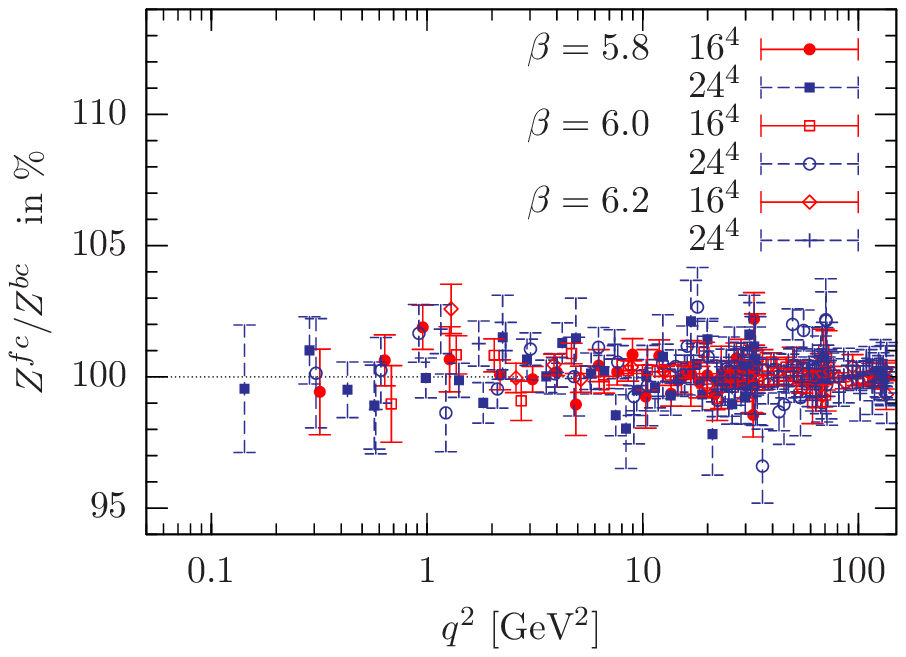} 
\hspace*{2.0cm}
\includegraphics[width=7.0cm
]{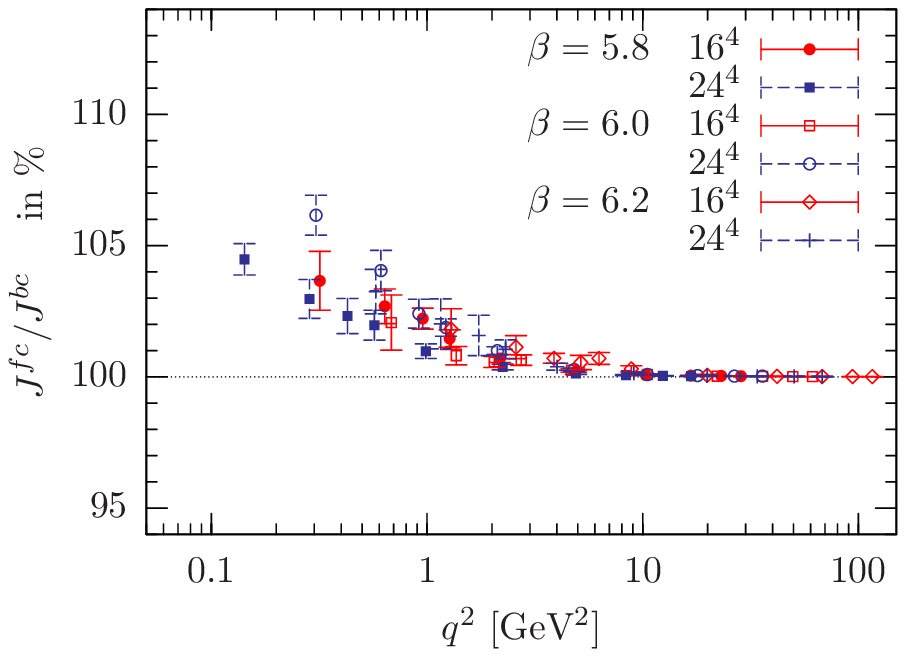}
\caption{
The ratios $Z^{\rm fc} / Z^{\rm bc}$ for the gluon dressing functions
(l.h.s.) and $J^{\rm fc} / J^{\rm bc}$ for the ghost dressing functions
(r.h.s) determined on first (\fc{}) and best (\bc{}) gauge copies,
respectively, as function of the momentum $q^2$.
}
\label{fig:fcbc_ratios}
\end{center}
\end{figure*}

Finally, in \Fig{fig:fcbc_ratios} we illustrate the effect of the 
Gribov copies for periodic gauge transformations.  
We have plotted some \fc{} -- to -- \bc{} ratios of the ghost and
gluon dressing functions. 
For the gluon propagator there is no influence revealing itself 
on top of the statistical noise. On the contrary, for the ghost propagator the 
Gribov problem can cause $O(5\%)$ deviations in the low momentum region 
($q<1~\mbox{GeV}$). For better gauge copies the ghost dressing function 
becomes less singular in the infrared. A closer inspection of the data 
for the ghost propagator indicates that the influence of Gribov copies 
becomes weaker for {\it increasing} physical volume. 
This probably corroborates a recent claim by Zwanziger 
telling that in the infinite volume 
limit averaging over gauge copies in the Gribov region should lead to 
the same result as averaging over copies restricted to the fundamental modular 
region \cite{Zwanziger:2003cf}. 
In \cite{Bogolubsky:2005wf} we have investigated for $SU(2)$ gluodynamics 
the effect of admitting a non-periodic extension (modulo $Z(2)$ flips) 
of the usually periodic gauge transformations. Also in this case we found that 
the effect on the Greens functions of the bias towards a particular gauge copy 
fades away with increasing lattice volume.
Anyway, we see that Gribov copy effects have to be studied properly in the 
infrared limit before one can come to final conclusions.

\section{Summary}
\label{sec:concl}
\vspace*{-0.2cm}
We have studied the low momentum region of QCD in the Landau gauge
using Monte Carlo simulations with the Wilson plaquette action. For
the quenched case we could afford to simulate on lattice sizes ranging
from $8^4$ up to $56^4$ using bare couplings constants $\beta$ in the 
interval $[5.7,\ldots,6.2]$. In this way we have presently reached 
momenta down to \mbox{$q\simeq 100$~MeV}. 
In order to assess the effect of virtual quarks we have carried out an
analogous investigation for full QCD with two flavors of 
clover-improved dynamical Wilson fermions for three different quark 
masses. For the time being we have evaluated only lattices of 
size $24^3 \times 48$. 
Our data presented here refer to an arbitrary (the ``first'') gauge copy
as obtained from over-relaxation or Fourier accelerated gauge fixing. 

Towards the infrared momentum region, the gluon dressing functions 
in quenched as well as in full QCD were shown to decrease, while the 
ghost dressing functions turned out to rise. However, the interrelated 
power laws predicted by the infinite-volume DSE approach could not 
(yet) be confirmed on the basis of our data. 
Our new data in the quenched case obtained for lattice sizes $48^4$ and 
$56^4$ at $\beta=5.7$ demonstrate that the gluon propagator flattens 
for $q^2 < 0.1 ~\textrm{GeV}^2$ leaving open the possibility for
its decrease at even lower momenta.

From the present data, the running coupling $\alpha_s(q^2)$ in the 
momentum subtraction scheme (based on the ghost-gluon vertex) does 
not seem to approach the expected finite infrared
fixed point monotonously.  It was rather seen to decrease for lower momenta 
after passing a turnover at $q^2 \simeq 0.3$~GeV$^2$ for 
quenched ($N_f=0$) as well as for full QCD ($N_f=2$).

Unquenching effects have been clearly identified for the gluon propagator,
whereas the ghost propagator was almost unchanged. This is in one-to-one
correspondence with what has been found in the Dyson-Schwinger 
equation approach. However, the puzzle of the existence of a non-trivial
infrared fixed point in the infinite volume limit remains unsolved. 

We have studied lattice effects as far as the dependence on the 
finite lattice size, on the lattice spacing and on the lattice asymmetry
are concerned. 
Within the parameter range under study the first two problems seem
to be under control, whereas infinite volume extrapolations based on the 
strong dependence on the lattice asymmetry might be worth to be further 
studied. 
 
Concerning the effect of Gribov copies we have seen a quite strong 
influence in the infrared region on the ghost propagator which becomes 
less singular when better gauge copies are taken. We have found some 
indications that the Gribov effect weakens as the volume increases.
\par\hfill

\vspace*{-0.5cm}
\section*{ACKNOWLEDGMENTS}
\vspace*{-0.2cm}
All simulations were done on the IBM pSeries 690 at HLRN and on
the MVS-15000BM at the Joint Supercomputer Center (JSCC) in Moscow.
This work was supported by the DFG under the contract FOR 465
(Forschergruppe Lattice Hadron Phenomenology), by the DFG-funded
graduate school GK~271 and with joint grants DFG 436 RUS 113/866/0 
and RFBR 06-02-04014. 
We thank the QCDSF collaboration for providing 
us their unquenched configurations which we could access within 
the framework of the I3 Hadron-Physics Initiative (EU contract 
RII3-CT-2004-506078). We are grateful to Hinnerk St\"uben for 
contributing parts of the program code. We also acknowledge useful
discussions with R.\ Alkofer, A.\ Cucchieri, G.\ Burgio, C.\ Fischer,
A.\ Maas, T.\ Mendes, V.\ K.\ Mitrjushkin, O.\ Oliviera, P.\ Silva, and
D.\ Zwanziger. M.\ M.-P.\ expresses his gratitude to the organizers of
`Infrared QCD in Rio' for the kind hospitality and the fruitful
workshop atmosphere. 

\bibliographystyle{apsrev}
\bibliography{references}

\end{document}